\documentstyle[prl,aps]{revtex}
\begin{document}
\twocolumn
\draft

\title{Teleportation, Bell's inequalities
and inseparability}

\author{Ryszard Horodecki \cite{mail}}

\address{Institute of Theoretical Physics and Astrophysics\\
University of Gda\'nsk, 80--952 Gda\'nsk, Poland}

\author{Micha\l{} Horodecki}

\address{Department of Mathematics and Physics\\
 University of Gda\'nsk, 80--952 Gda\'nsk, Poland}

\author{Pawe\l{} Horodecki}

\address{Faculty of Applied Physics and Mathematics\\
Technical University of Gda\'nsk, 80--952 Gda\'nsk, Poland}

\maketitle

\begin{abstract}
Relations  between teleportation, Bell's inequalities and inseparability 
are investigated. It is  shown that any mixed two spin-$1\over2$ state 
which violates the Bell-CHSH inequality is useful for teleportation.
The result is  extended to any Bell's inequalities constructed of the 
expectation values of products of spin operators. It is also shown that
there exist inseparable states which  are not useful for teleportation
within the standard scheme.
\end{abstract}
%\newpage

Recently Bennett at al. \cite{bennett} have discovered a new
aspect of quantum inseparability -- teleportation.
It involves a separation of an input state into classical and quantum part
from which the state can be reconstructed with perfect fidelity ${\cal F}=1$.
The basic idea is to use a pair of particles in singlet state shared by
sender (Alice) and receiver (Bob).
Quite recently Popescu \cite{pop1} noticed that the pairs in
a mixed state  could be still useful for (imperfect) teleportation. There was
a question what value of fidelity of the transmission of an unknown state can
ensure us about nonclassical character of the state forming the quantum
channel. It has been shown \cite{pop1,massar} that the purely classical channel
can give at most ${\cal F}={2\over3}$
(see also Ref. \cite{Gis} in this context). Then Popescu raised basic 
questions concerning a possible relation between teleportation, Bell's
inequalities and inseparability:
``What is the exact relation between Bell's inequalities
violation and teleportation?
Is every mixed state that can not be expressed as a
mixture of product states useful for teleportation?'' \cite{pop1} 
\footnote{We call a state inseparable if it cannot be written as convex 
combination of product states}.

The problem is rather complicated, as these questions concern the mixed states
which apparently possess the ability to behave classically in some respect 
but quantum mechanically in others \cite{pop1}.
Fortunately for $2\times2$ systems two basic questions concerning violation
of Bell's inequalities and inseparability of mixed states, have been solved 
completely. In particular, in Ref.\cite{my} the effective criterion for 
violation of Bell's inequalities has been  obtained. Quite recently 
the problem of inseparability have been investigated in detail by 
Peres \cite{Peres} and the authors \cite{sep}. In particular, the necessary
\cite{Peres} and sufficient \cite{sep} condition for separability of mixed 
states for $2\times2$ systems has been provided. 

The main purpose of the present Letter is to present the effective
criterion for teleportation via mixed two spin-$1\over2$ states and 
discuss it in the context of Bell's inequalities and inseparability. 
Using the results contained in Refs. \cite{my,Peres,sep}
we will show further that if a mixed two spin-$1\over2$  state violates 
any in Bell's inequality constructed of the expectation values of 
products of spin operators (in particular if it violates original 
Bell-CHSH one), then it is also useful for teleportation.  We will also
demonstrate that there are inseparable states which are not useful for 
teleportation within the standard scheme \footnote{By the standard 
teleportation scheme we mean here that Alice uses Bell operator basis 
\cite{Mann} in her measurement while Bob is allowed to apply any unitary 
transformation.}.

\section{Maximal fidelity for the standard teleportation scheme}

We start  with the representation of the state in the
Hilbert-Schmidt space
\begin{equation}
\varrho={1\over 4}[{I}\otimes{I}+\bbox{r\cdot\sigma}\otimes{I}+
 {I}\otimes\bbox{s\cdot\sigma}+\sum^3_{n,m=1}t_{nm}\sigma_n\otimes\sigma_m]
 \label{hs}
\end{equation}
where $\varrho$ acts on Hilbert space
${\cal H}={\cal H}_1\otimes{\cal H}_2=C^2\otimes C^2$,
${I}$ stands for identity operator, $\{\sigma_n\}^3_{n=1}$ are the
standard Pauli matrices, $\bbox r, \bbox s$  are  vectors in ${R^3}$,
$\bbox{r\cdot\sigma}=\sum_{i=1}^3 r_i \sigma_i$. The coefficients
$t_{nm}=\text{Tr} (\varrho\sigma_n\otimes\sigma_m)$ form a real matrix which we
shall denote by $T$.
Note that the  representation appears to be very convenient in the
investigation of some aspects of inseparability of the mixed states.
Indeed, all the parameters fall into two different classes: first
($\bbox{r}$ and $\bbox{s}$) --
describing the local behaviour of the state, second ($T$ matrix) --
responsible for correlations.
It is compatible with the fact that the mean value
of the Bell-CHSH observable depends  only on the correlation parameters $T$
of the state $\varrho$ \cite{my}.

Let us briefly recall the standard teleportation scheme.
It involves two
particle source producing pairs in a given mixed state $\varrho$ which forms
the quantum channel (originally formed by pure singlet state
\cite{bennett}). One of the particles is given to Bob while the other one
and a third particle in an unknown state $\phi$ are subjected to
Alice's joint measurement. The latter is given by a family of projectors
\begin{equation}
P_k=|\psi_k\rangle\langle\psi_k|
\quad k=0,1,2,3,
\label{singlety}
\end{equation}
where  $\psi_k$ constitute the so-called Bell basis
\begin{eqnarray}
&&\psi_{1\atop(2)}={1\over\sqrt2}(e_1\otimes e_1\mp e_2\otimes e_2)\nonumber\\
&&\psi_{3\atop(0)}={1\over\sqrt2}(e_1\otimes e_2\pm e_2\otimes e_1)
\label{baza}
\end{eqnarray}
with $e_1,e_2$ being standard basis in $C^2$.
Then using two bits Alice sends
to Bob the number of outcome $k$ and Bob applies some unitary transformation
$U_k$
obtaining in this way his particle in a
state $\varrho_k$. 

Then the fidelity of a transmission of the unknown state
is given by formula \cite{pop1,Gis,inf}
\begin{equation}
{\cal F}=\int_S\text dM(\phi)\sum_kp_k\text{Tr}(\varrho_k P_\phi)
\label{fidelity}
\end{equation}
where the integral is taken over all $\phi$ belonging to the Bloch sphere
with uniform distribution $M$,
$p_k=\text{Tr}\left[(P_k\otimes I)(P_{\phi}\otimes
\varrho)\right]$ denotes the probability of the $k$-th outcome. Now the task
is to find such $U_k$'s that produce the highest fidelity (a choice of a
quadruple of $U_k$'s we shall call strategy).
In this purpose, let us compute the integral (\ref{fidelity}).
The output state  $\varrho_k$ is
given by
\begin{equation}
\varrho_k={1\over p_k}\text{Tr}_{1,2}\left[(P_k\otimes U_k)
(P_{\phi}\otimes\varrho) (P_k\otimes U_k^\dagger)\right]
\label{output}
\end{equation}
Here the partial trace is taken over the states of unknown particle and
Alice's one.
Putting $P_\phi={1\over2}(I+\bbox a\cdot \bbox \sigma)$ one obtains
\begin{equation}
p_k\varrho_k={1\over8}([1+(\bbox a,T_k\bbox r)]I+
O_k^\dagger[\bbox s+T^\dagger T_k\bbox a]\cdot\bbox\sigma)
\end{equation}
Here $T_k$'s and $\bbox r$, $\bbox s$, $T$ correspond to $P_k$'s and
$\varrho$ respectively via formula (\ref{hs})
(we have: $T_0=\text{diag}(-1,-1,-1)$,  $T_1=\text{diag}(-1,1,1)$
$T_2=\text{diag}(1,-1,1)$, $T_3=\text{diag}(1,1,-1)$,
$\bbox r_k=\bbox s_k=0$, for $k=0,1,2,3$.);
$O_k$'s are rotations  in $R^3$ obtained from $U_k$'s by
\begin{equation}
U\bbox{\hat n\cdot\sigma}U^\dagger
=(O^\dagger\bbox{\hat n})
\bbox{\cdot\sigma},
\label{oplusy}
\end{equation}
($O$ is here determined uniquely as the group of rotations $O^+(3)$ is a
homomorphic image of U(2) group \cite{wies}).
Omitting the terms which do not contribute to the integral (\ref{fidelity})
and using the formula
\begin{equation}
\int_S(\bbox a,A\bbox a)\text d M(\bbox a)
={1\over3}\text{Tr} A
\end{equation}
one obtains
\begin{equation}
{\cal F}={1\over8}\sum_k(1+{1\over3}\text{Tr}T_k^\dagger TO_k)
\end{equation}

Now we shall maximize $\cal F$ under all strategies.
Clearly, as $-T_k^\dagger$ is a rotation 
we see that the
maxima of the terms in the above formula do not depend on $k$ any longer so
that 
\begin{equation}
{\cal F}_{\max} =\max_O{1\over2}(1-{1\over3}\text{Tr}TO)
\label{fidelity2}
\end{equation}
where the maximum is taken over all rotations.
Note that we have
\begin{equation}
{\cal F}_{\max}\leq {1\over2}(1+{1\over3}\text{Tr}\sqrt{T^\dagger T})
\label{granica}
\end{equation}
Of course, we need to derive the expression for ${\cal F}_{\max}$
only
if the latter is greater than $2\over3$ which is the upper bound for the
classical
teleportation \cite{pop1,massar}. If ${\cal F}_{\max}>{2\over3}$ we say that the
state forming the quantum channel is useful for teleportation. Clearly, ${\cal
F}_{\max}$ can exceed ${2\over3}$ only if $\text{Tr}\sqrt{T^\dagger T}>1$.
Now basing on the results contained in Ref. \cite{inf} one can see that  
the latter condition implies $\det T<0$. But then
the inequality (\ref{granica}) passes into equality.
Consequently, defining function $N(\varrho):=\text{Tr}\sqrt{T^\dagger T}$
one has

\noindent
{\bf Proposition 1} {\it
Any mixed spin-$1\over2$ state is useful for (standard) teleportation iff
$N(\varrho)>1$. Then the fidelity amounts to
\begin{equation}
{\cal F}_{\max}={1\over2}(1+{1\over3}N(\varrho)).
\end{equation} }

\noindent
Now, if $N(\varrho)>1$ then
there exist rotations $O_1$
and $O_2$ such that $O_1TO_2$ is diagonal with $t_{ii}<0$ for $i=1,2,3$. Then
the best strategy is given by unitaries $U_k=U\sigma_k$ where $U$ is determined
(up to an irrelevant phase factor) by $O=O_1O_2$ via formula (\ref{oplusy}).

{\it Example.-} Consider pure state of the form 
\begin{equation}
|\psi\rangle=a e_1\otimes e_2- be_2\otimes e_1.
\end{equation}
One obtains 
\begin{equation}
{\cal F}_{\max}={2\over3} {a^3-b^3\over a-b},
\end{equation}
which is compatible with Ref. \cite{Gis}.

\section{Relation between Bell's inequalities and teleportation}
Note that $N(\varrho)$ is a function of the correlation parameters $T$
only. It allows to establish a relation between teleportation and Bell's
inequality due to Clauser, Horne, Shimony and Holt \cite{chsh} (Bell-CHSH).
As one knows the necessary and sufficient
condition for violating the Bell-CHSH inequality involves a real valued
function $M(\varrho)=\max_{i>j}(u_i+u_j)$ where $u_i$ are eigenvalues of
matrix $T^\dagger T$ \cite{my}. Then the inequality $M(\varrho)\leq1$ is
equivalent to the Bell-CHSH one. Now as $u_i\leq1$ for $i=1,2,3$ 
\cite{cor,myprl}
and $N(\varrho)=\sum_{i=1}^3\sqrt{u_i}$
we obtain a relation
\begin{equation}
N(\varrho)\geq M(\varrho).
\label{nm}
\end{equation}
Note that for any state which violates the Bell-CHSH inequality
we have $M(\varrho)>1$.
Then, according to the relation (\ref{nm}) and Prop. 1 we get
the estimate
\begin{equation}
{\cal F}_{\max}
\geq{1\over2}(1+{1\over3}M(\varrho))>{2\over3}.
\label{niera}
\end{equation}
As the maximal mean value of the CHSH-Bell observable is
$B_{max}=2\sqrt{M(\varrho)}$ we have also
\begin{equation}
{\cal F}_{\max}
\geq{1\over2}(1+{1\over12}B^2_{max}).
\label{nierb}
\end{equation}
The inequalities (\ref{niera}), (\ref{nierb}) are valid for an arbitrary mixed
two spin-$1\over2$ state which violate the Bell-CHSH inequality and they say us
that {\it any} such a state is useful for teleportation.

Now we shall see that even a stronger statement is valid.
For this purpose consider generalized Bell-CHSH inequalities i.e. all the
Bell's inequalities which can
be constructed of the expectations of products of spin operators
$\bbox{a\cdot\sigma} \otimes
\bbox{b\cdot\sigma}$ where $\bbox{a}$ and $\bbox{b}$ are unit vectors
\cite{roy}.
Of course, the expectations (or correlation functions)
\begin{equation}
E(\bbox{a},\bbox{b})
\equiv \text{Tr}(\varrho\bbox{a\cdot\sigma}\otimes\bbox{b\cdot\sigma})=(\bbox
a,T\bbox b)
\end{equation}
depend only on the $T$ matrix.
Hence the generalized
Bell-CHSH inequalities can be violated only if $N(\varrho)>1$.
Indeed, if $N(\varrho)\leq1$, there always exists some
separable state
that has the same $T$ matrix  as the state
$\varrho$ (see Ref. \cite{inf}).
In this way we have obtained

\noindent
{\bf Proposition 2} {\it
Every mixed two spin-$1\over2$ state which violates any generalized Bell-CHSH
inequality is useful for teleportation.}

\section{Inseparability and teleportation}
Let us now turn back to  the question: ``Is every mixed state that can not
be expressed as a mixture of product states useful for teleportation?''.
Generally, the problem is rather complicated as it requires to obtain the
maximum of the fidelity over all possible teleportation procedures.
Here we will see that within the standard teleportation scheme the
answer is ``no''. For this purpose consider the following class of the states
\begin{equation}
\varrho=p_1|\psi_1\rangle\langle\psi_1|+p_2|\psi_2\rangle\langle\psi_2|
\label{stany}
\end{equation}
where
\begin{eqnarray}
|\psi_1\rangle =ae_1\otimes e_1+be_2\otimes e_2\\
|\psi_2\rangle =ae_1\otimes e_2+be_2\otimes e_1
\label{czyste}
\end{eqnarray}
with $a,b>0$, $\{e_i\}$ being standard basis in $C^2$,
\ $0<(p_1-p_2)^2\leq(a^2-b^2)^2$.
The above states have interesting properties.
First, note that as $M(\varrho)=1+(p_1-p_2)^2-(a^2-b^2)^2\leq1$ they do not
violate the Bell-CHSH inequality. In addition, it is possible to choose the
parameters $p_1$ and $a$ so that the maximal absolute value of the
expectation of products of spin operators
is arbitrarily close to zero. Then it follows from
Prop. 1 that many of the states (\ref{stany}) are not useful for teleportation.
But what can we say about the above states in the context of the
inseparability ? As it was mentioned in the introduction, the effective
criterion for inseparability of the states of $2\times 2$ systems
have been found \cite{Peres,sep}. Namely a two spin-$1\over2$ state is 
inseparable if and only if its partial transposition is not a positive 
operator. The matrix elements of partial transposition
$\varrho^{T_2}$ of a state $\varrho$ is given by 
\begin{equation}
\varrho^{T_2}_{m\mu,n\nu}\equiv\varrho_{m\nu,n\mu},
\end{equation}
 where 
\begin{equation}
\varrho_{m\mu,n\nu}=\langle e_m\otimes e_\mu |\varrho|e_n\otimes f_\nu
\rangle. 
\end{equation}
Now  is easy to see that all the states \ref{stany} 
are inseparable \footnote{In Ref. \cite{fazy} the states \ref{stany}
were shown to be inseparable by means of entropic criterion (see in this 
context  Ref. \cite{inf}. 
 The latter
appears to be equivalent to inseparability for the considered states, but 
it is not the case in general \cite{Peres,sep}.}.
In fact, one can show
that the inseparability of the above states  manifests itself
via ``hidden'' nonlocality (see Ref. \cite{pop2}) which can be revealed
\cite{fazy} by means of Gisin's filtering method \cite{gisin}.
Thus we have provided an example of  states which are inseparable and
nonlocal but still are not useful for teleportation
within the standard scheme.

\section{Conclusion}

In conclusion, we have considered the questions concerning possible relations
between teleportation, violation of Bell's inequalities and
inseparability. In particular, we have obtained the maximal fidelity for the
standard teleportation scheme with the quantum channel formed by any mixed
two spin-$1\over2$ state. It involves only the correlation parameters of the
state. Then
it was possible to compare the two
different aspects of quantum inseparability: teleportation and Bell's
inequalities. More precisely, we have shown that
if a mixed two spin-$1\over2$
state violates any generalized Bell-CHSH inequality (in particular  if it violates
the original Bell-CHSH one) then it is also useful for
teleportation.

We have also considered the states which are inseparable, but are not
useful for the standard teleportation. Here the inseparability is due to the
relation between the local and correlation parameters.
Then there is a question: what would happen if we allowed Alice to use
any projectors -- not only the maximally entangled ones? %\cite{maxy}? 
In
fact she may perform any generalized measurements. In the formula
for the fidelity the local parameters could then also appear. It is not clear
whether a higher fidelity can be obtained within so generalized
scheme. Thus, the problem of a relation between the
widely understood Bell's inequalities (e.g. involving nonstandard measurements
\cite{pop1,fazy,pop2,gisin}) and more general teleportation schemes needs 
further investigations.

We would like to acknowledge stimulating discussions with Nicolas
Gisin.

\end{document}